# Efficient factored gradient descent algorithm for quantum state tomography

Yong Wang,[1] Lijun Liu,[2,*] Shuming Cheng,[1,3,4,†] Li Li,[1,3] and Jie Chen[1,3]
[1]College of Electronics and Information Engineering, Tongji University, Shanghai 201804, China
[2]Department of Mathematics and Computer Science, Shanxi Normal University, Taiyuan 030006, China
[3]Shanghai Research Institute for Intelligent Autonomous Systems, Tongji University, Shanghai 201203, China
[4]Institute for Advanced Study, Tongji University, Shanghai 200092, China

(Received 6 November 2023; accepted 20 June 2024; published 8 July 2024)

Reconstructing the state of quantum many-body systems is of fundamental importance in quantum information tasks, but extremely challenging due to the curse of dimensionality. In this work, we present an efficient quantum tomography protocol that combines the state-factored with eigenvalue mapping to address the rank-deficient issue and incorporates a momentum-accelerated gradient descent algorithm to speed up the optimization process. We implement extensive numerical experiments to demonstrate that our factored gradient descent algorithm efficiently mitigates the rank-deficient problem and admits orders of magnitude better tomography accuracy and faster convergence. We also find that our method can accomplish the full-state tomography of random 11-qubit mixed states within one minute.

## I. INTRODUCTION

Quantum state tomography (QST) is a powerful tool to recover full-state information of the unknown many-body quantum systems from measurement statistics, and thus plays an indispensable role in quantum information processing tasks, with wide applications ranging from certifying fundamental principles in quantum theory [1,2] and benchmarking quantum devices [3,4], to verifying quantum algorithms [5,6]. However, it is a challenging task due to the curse of dimensionality in the sense that it admits an exponential growth of measurement settings, memory cost, and computing resources as the number of subsystems involved linearly increases [4].

Typically, the way QST works is that some proper state estimator, such as the maximum-likelihood estimator (MLE) [7–10] or linear regression estimator (LRE) [11–13], is introduced to be optimized over parameterized estimate states and to find a (sub)optimal state close to the target state [14]. Several efficient QST methods have been developed for specific classes of states, including low-rank states [15–17], permutationally invariant states [18,19], matrix product states [20–24], and neural states [25–29]. Since a physical state must satisfy the fundamental constraint of positivity, extra techniques are needed to map the estimated state to a physical one. For example, one is to parametrize the state via the positive-semidefinite (PSD) factorization [7,9,16,17,28,30–32], and another is state projection which maps eigenvalues (possibly negative) of the estimate to those of physical states [11,13,33–37]. However, the rank-deficient issue exists in those estimators and is likely to output the estimate states with at least one or more zero eigenvalues [9,11–13,38,39] for the full-rank target states. Even though the estimators such as the hedged MLE [38] and Bayesian mean estimator [36,39] are able to yield the almost full-rank state, they usually require heavy computation. Thus, it is crucial to design a state estimator which mitigates the rank-deficit problem and can also efficiently find a physical estimate.

Further, numerous gradient-based algorithms for the factored or projected methods have been employed to accelerate the QST convergence, with some well-known examples including diluted direct gradient [8], conjugate gradient (CG) [40], projected gradient (PG) [35], CG-accelerated PG (CG-APG) [40], momentum-inspired gradient descent (MGD) [41], stochastic gradient [42], and Riemannian gradient [43]. However, the issue exists that the current factored gradient methods have limited convergence speed and accuracy, and the projected ones need extra computational resources [40].

To address the above issues, here we propose a state-mapping protocol based on factorization and $P$-order absolute eigenvalue mapping $\mathcal{A}_P$, which mitigates the rank-deficient problem in the tomography optimization. We further show that this technique is also capable to leverage other QST methods, such as the MLE with MGD [41]. Additionally, we introduce a momentum-accelerated Rprop (MRprop) gradient descent algorithm to achieve fast accurate MLE reconstruction. Moreover, a product-structured positive operator-valued measure (POVM) is employed to measure the system and to reduce the storage cost as well as the required computational resources.

Extensive numerical experiments are implemented on a large number of random states, ranging from pure states and low-rank states, to maximally entangled states. Our results demonstrate that the proposed factored gradient descent algorithm significantly improves the tomography speed and accuracy of MLE on states with various purity and rank. This improvement is also witnessed in the MLE using other optimizer MGD [41] and the LRE [12]. Furthermore, it is found that the MLE with MRprop and $\mathcal{A}_P$ mapping achieves the optimal infidelity scale of $O(1/N)$ with sample size $N$ [10] and a full-state tomography of random 11-qubit mixed states within one minute, while the MLE with CG-APG takes nearly 15 minutes for 10-qubit states [40]. Finally, the results also show that our scheme outperforms the iterative MLE (iMLE) [44] and MLE with CG-APG by orders of magnitude less runtime and iterations, and more accuracy than the LRE.

*Contact author: lljcelia@126.com
†Contact author: shuming_cheng@tongji.edu.cn

The remainder of this paper is organized as follows. Section II introduces basic concepts related to QST, including state description, MLE and LRE estimators, state-mapping techniques, and state fidelity. Section III details our state-mapping method, MRprop gradient descent algorithm, and product-structured POVM. Then, the numerical results and analysis are presented in Sec. IV. Finally, the summary and outlook are discussed in Sec. V.

## II. QUANTUM STATE TOMOGRAPHY

### A. Reconstructing the density matrix

The state of quantum many-body systems is characterized by a PSD density matrix $\rho$ with unit trace, i.e., $\rho \succeq 0$ and $\mathrm{Tr}[\rho] = 1$. Formally, it can be written as

$$\rho = \sum_i p_i |\psi_i\rangle\langle\psi_i|, \tag{1}$$

where $p_i$ denotes the occurrence probability of state $|\psi_i\rangle$ satisfying $p_i \geqslant 0$ and $\sum_i p_i = 1$. If $\rho = |\psi\rangle\langle\psi|$, then $\rho$ is pure, otherwise it is a mixed state. Note that for the $n$-qubit system, generically, it requires $d^2 - 1$ real parameters to fully determine $\rho$, where $d = 2^n$ is the dimension of the Hilbert space. It is obvious that the number of parameters describing state $\rho$ grows exponentially with the qubit number $n$.

To reconstruct density matrix $\rho$ in Eq. (1), the QST is decomposed of two steps: The first is the measurement procedure which yields the outcome statistics described by a probability distribution (PD) generated from measuring on identically prepared copies of unknown $\rho$. Specifically, quantum measurement is modeled as a POVM $\{M_k\}_{k=1}^K$ with $M_k \succeq 0$ and $\sum_{k=1}^K M_k = \mathbb{1}$ ($\mathbb{1}$ is the identity matrix), and the probability $p_k$ of each outcome $k$ is governed via the Born rule with $p_k = \mathrm{Tr}[M_k\rho]$. The second is the reconstruction procedure which finds an estimate state $\tilde{\rho}$ from the observed PDs. It can be formulated as the following optimization problem:

$$\begin{aligned} &\text{estimate} && \tilde{\rho} \\ &\text{subject to} && \tilde{\rho} \succeq 0 \text{ and } \mathrm{Tr}[\tilde{\rho}] = 1, \\ &&& p_k = \mathrm{Tr}[M_k\rho] = \mathrm{Tr}[M_k\tilde{\rho}] \ \ \forall M_k, \\ &&& M_k \succeq 0 \text{ and } \sum_{k=1}^K M_k = \mathbb{1}. \end{aligned} \tag{2}$$

In practice, it is challenging to directly solve the above problem for the systems with a larger number of qubits, due to the curse of dimensionality that as qubit number $n$ linearly increases, it requires an exponential growth of measurement settings $\{M_k\}_{k=1}^K$ to determine $\rho$, memory to store $\boldsymbol{p} = (p_1, \ldots, p_K)$, and computational resources to process the reconstruction. Further, the presence of statistical error and readout noise leads to the noisy data $f_k \approx \mathrm{Tr}[M_k\rho]$, instead of the accurate probability $p_k$, which makes the tomography task more challenging.

### B. The state estimators

#### 1. Maximum-likelihood estimator

One commonly used choice to perform state estimation is MLE that minimizes the negative log-likelihood function between the measured frequency $f_k$ and the estimate probability $\tilde{p}_k = \mathrm{Tr}[M_k\tilde{\rho}]$ [7–9],

$$\begin{aligned} &\underset{\boldsymbol{\theta}}{\text{minimize}} && -\sum_k f_k \ln(\mathrm{Tr}[M_k\mathcal{P}[T(\boldsymbol{\theta})]]), \\ &\text{subject to} && M_k \succeq 0 \text{ and } \sum_{k=1}^K M_k = \mathbb{1}, \end{aligned} \tag{3}$$

where $\tilde{\rho} = \mathcal{P}[T(\boldsymbol{\theta})]$ is a physical estimate state with a proper state mapping $\mathcal{P}[\cdot]$, and $T(\boldsymbol{\theta})$ represents the estimate matrix characterized by parameter vector $\boldsymbol{\theta}$. It has a unique global minimum for strictly convex negative log-likelihood functions in Eq. (3), otherwise some local minima may exist far from the global one [45]. One of the well-known MLE algorithms is iMLE, which iteratively constructs correction operators to make the state $\tilde{\rho}$ asymptotically approach the maximum-likelihood estimate, via the expectation maximization [44]. The MLE has the advantage of asymptotic normality and efficiency for full-rank states [10,45], and has been tested experimentally on, e.g., the single-ion Zeeman qubit [46] and entangled photons [47].

#### 2. Linear regression estimator

With a complete Hermitian operator basis set $\{\Omega_i\}_{i=1}^{d^2}$, any state can be characterized by parameter vector $\boldsymbol{\theta} = (1/\sqrt{d}, \theta_2, \ldots, \theta_{d^2})^\mathsf{T}$ via

$$\rho = \sum_{i=1}^{d^2} \theta_i \Omega_i, \tag{4}$$

where $\Omega_1 = \mathbb{1}/\sqrt{d}$, $\mathrm{Tr}[\Omega_i] = 0$ for $i = 2, \ldots, d^2$, and $\mathrm{Tr}[\Omega_i\Omega_j] = \delta_{ij}$ with the Kronecker function $\delta_{ij}$. Similarly, the POVM $\{M_k\}_{k=1}^K$ is also decomposed into $M_k = \sum_{i=1}^{d^2} X_{ki}\Omega_i$ for $k = 1, \ldots, K$, and $X$ is a representation matrix with elements $X_{ki} = \mathrm{Tr}[M_k\Omega_i]$. Thus, the measurement possibility $p_k$ associated with outcome $k$ is determined by $p_k = \mathrm{Tr}[M_k\rho] = \sum_{i=1}^{d^2} X_{ki}\theta_i = X_k\boldsymbol{\theta}$, with $X_k = (X_{k1}, \ldots, X_{kd^2})$.

Another state estimator is LRE [11–13] formulated as the optimization problem

$$\begin{aligned} &\underset{\boldsymbol{\theta}}{\text{minimize}} && \sum_k (f_k - X_k\boldsymbol{\theta})^2, \\ &\text{subject to} && M_k \succeq 0 \text{ and } \sum_{k=1}^K M_k = \mathbb{1}. \end{aligned} \tag{5}$$

It admits a closed-form solution $\hat{\boldsymbol{\theta}} = (X^\mathsf{T}X)^{-1}X^\mathsf{T}\boldsymbol{f}$, where $\boldsymbol{f} = (f_1, \dots, f_K)^\mathsf{T}$ denotes the frequency vector and $A^\mathsf{T}$ is the transpose of matrix or vector $A$. Compared to MLE, the LRE can offer faster arithmetic but slightly worse accuracy, with experimental validation reported in [11].

### C. Mapping techniques to ensure the positivity of estimate states

In the state estimators of MLE (3) and LRE (5), the estimate state $\tilde{\rho}$ should satisfy the fundamental physical constraint that it is a PSD operator with unit trace; however, this may not always be guaranteed. As a consequence, it is essential to introduce the state mapping $\mathcal{P}[\cdot]$ of either state factorization or projection to map the unphysical estimate $T$ to physical state $\tilde{\rho}$.

#### 1. Factorization method

An arbitrary density matrix $\tilde{\rho}$ admits the factorization

$$\tilde{\rho} = \mathcal{P}[T] = \frac{T^\dagger T}{\mathrm{Tr}[T^\dagger T]}, \tag{6}$$

where transition matrix $T$ is usually a complex lower triangular matrix (Cholesky factorization) [7,28,30–32] or an arbitrary complex matrix [9,16,17], and † denotes the complex conjugate transpose.

It is evident from Eq. (6) that $\tilde{\rho}$ is automatically positive, so the QST optimization is reduced to find $T$ in the factored space without the positivity constraint. There is evidence to show the suboptimal tomography accuracy of the existing factored methods [40].

#### 2. Projection method

Note that any Hermitian matrix $T$ has the spectral decomposition $T = \mathcal{V}\Lambda\mathcal{V}^\dagger$, where $\Lambda$ is the eigenvalues diagonal matrix with eigenvalues $\boldsymbol{\lambda}$, and $\mathcal{V}$ is the eigenvectors unitary matrix. The state-projection methods [11,13,33–37] first introduce the mapping $\mathcal{P}_\triangle(\cdot)$, which maps the eigenvalues $\boldsymbol{\lambda}$ (possibly negatives) to vector $\boldsymbol{\sigma}$ on the unit simplex $\triangle$ under certain distance function $f$, i.e.,

$$\boldsymbol{\sigma} = \mathcal{P}_\triangle(\boldsymbol{\lambda}) = \arg\min_{\boldsymbol{\sigma}\in\triangle} f(\boldsymbol{\sigma}, \boldsymbol{\lambda}), \tag{7}$$

with

$$\triangle := \{\boldsymbol{x} \in \mathbb{R}^d \mid \boldsymbol{x} \geqslant 0, \sum_{i=1}^{d} x_i = 1\}. \tag{8}$$

Then, the physical estimate state is constructed from the projected $\boldsymbol{\sigma}$ and $\mathcal{V}$,

$$\tilde{\rho} = \mathcal{P}_\triangle[T] \equiv \mathcal{V}\mathcal{P}_\triangle(\Lambda)\mathcal{V}^\dagger \equiv \mathcal{V}\Sigma\mathcal{V}^\dagger. \tag{9}$$

Here the diagonal matrix $\Sigma$ is formed by vector $\boldsymbol{\sigma}$.

There are many choices of distance $f$. For example, it is mostly chosen as the square of the Frobenius norm, $||\boldsymbol{\sigma} - \boldsymbol{\lambda}||_F^2 := (\boldsymbol{\sigma} - \boldsymbol{\lambda}) \cdot (\boldsymbol{\sigma} - \boldsymbol{\lambda})$, and the corresponding state projection is called the simplex projection $\mathcal{S}[\cdot]$ [34,35,37]. It has an efficient solution: (1) It sets $\lambda_i = \lambda_i + (1 - \sum_i \lambda_i)/a$ for all $\lambda_i$ and the element number $a$ of $\{\lambda_i\}$. (2) If all $\lambda_i \geqslant 0$, then $\sigma_i = \lambda_i$ for all $\lambda_i$ from $\{\lambda_i\}$, and end the search. Otherwise, set $\sigma_i = \lambda_i = 0$ for all $\lambda_i < 0$ from $\{\lambda_i\}$, remove $\lambda_i \leqslant 0$ from $\{\lambda_i\}$, and repeat step (1).

An alternate is to first project $\boldsymbol{\lambda}$ to the positive space $\mathrm{Re}^+$ and then to unit simplex $\triangle$ [48],

$$\boldsymbol{\sigma}' = \mathcal{P}_{\mathrm{Re}^+}(\boldsymbol{\lambda}) = \arg\min_{\boldsymbol{\sigma}'\in\mathrm{Re}^+} f(\boldsymbol{\sigma}', \boldsymbol{\lambda}), \tag{10}$$

with

$$\mathrm{Re}^+ := \{\boldsymbol{x} \in \mathbb{R}^d \mid \boldsymbol{x} \geqslant 0\}. \tag{11}$$

We can obtain a physical state $\tilde{\rho} = \mathcal{V}\Sigma\mathcal{V}^\dagger$ similar to Eq. (9), with $\boldsymbol{\sigma} = \mathcal{P}_\triangle(\boldsymbol{\sigma}')$. The distance $f$ in Eq. (10) could be either the square of the Frobenius norm or nuclear norm, which can be efficiently solved via $\boldsymbol{\sigma}' = \max(\boldsymbol{\lambda}, 0)$. This stepwise projection $\mathcal{P}_\triangle \circ \mathcal{P}_{\Re^+}[\cdot]$ is named $\mathcal{M}[\cdot]$ and will be examined in our experiments in Sec. IV. We remark that it can also be projected first to the unit sum space $\{\boldsymbol{x} \in \mathbb{R}^d \mid \sum_{i=1}^d x_i = 1\}$ using the Frobenius norm, and then to unit simplex, equivalent to the $\mathcal{S}$ projection.

If $T$ is non-Hermitian, $(T + T^\dagger)/2$ yields a Hermitian one that is processed with the above projection process. Indeed, the projection method pulls the estimate matrix $T$ close to a physical $\tilde{\rho}$ to enhance the tomography performance, with expensive spectral decomposition and projection operator [36,40,48]. Further, discarding negative eigenvalues in the $\mathcal{S}$ or $\mathcal{M}$ projection signals the possibility of state information loss and also gives rise to the estimate close to the lower-rank states for full-rank target states, thus raising the rank-deficient issue. Thus, it still calls for developing new state-mapping techniques to avoid such issues, whereas for the rank-deficient states, projection methods provide comparable estimates.

### D. State fidelity metric for tomography

The reconstruction accuracy of QST estimators can be evaluated via quantum state fidelity $F_q$, which measures the distance between the target state $\rho$ and the estimate $\tilde{\rho}$. Specifically, the fidelity $F_q$ is defined as [49,50]

$$F_q(\rho, \tilde{\rho}) = (\mathrm{Tr}[\sqrt{\sqrt{\tilde{\rho}}\rho\sqrt{\tilde{\rho}}}])^2, \tag{12}$$

and state infidelity is $1 - F_q(\rho, \tilde{\rho})$. An efficient approach exists to compute the state fidelity as [51,52]

$$\begin{aligned}\sqrt{F_q(\rho, \tilde{\rho})} &= \sum_i \lambda_i\Big(\sqrt{\sqrt{\tilde{\rho}}\rho\sqrt{\tilde{\rho}}}\Big) \\ &= \sum_i \sqrt{\lambda_i(\sqrt{\tilde{\rho}}\rho\sqrt{\tilde{\rho}})} = \sum_i \sqrt{\lambda_i(\rho\tilde{\rho})},\end{aligned} \tag{13}$$

where $\lambda_i(A)$ refer to the eigenvalues of matrix $A$, which are non-negative reals for the PSD matrices $\rho$ and $\tilde{\rho}$ [52]. Obviously, Eq. (13) only needs to find the eigenvalues of $\rho\tilde{\rho}$, rather than using the two computationally expensive matrix square roots in Eq. (12).

## III. FACTORED MRPROP ALGORITHM WITH $P$-ORDER ABSOLUTE STATE MAPPING

In this section, we first present a state-mapping method based on factorization and eigenvalue mapping, called $P$-order absolute state mapping $\mathcal{A}_P[\cdot]$, to alleviate the rank-deficient issue in MLE and to ensure the fast QST. Then, we develop the MRprop gradient algorithm to expedite the state reconstruction. We find that our MLE with MRprop and $\mathcal{A}_P$ mapping is capable of achieving the full-state tomography of 11-qubit random mixed states within one minute, while it takes nearly 15 minutes for 10-qubit states using the superfast MLE with CG-APG [40].

### A. *P*-order absolute state-mapping method

#### 1. Method design

For the Hermitian matrix $T$, it follows directly from the factored method that we have

$$\tilde{\rho} = \frac{T^\dagger T}{\mathrm{Tr}[T^\dagger T]} = \frac{\mathcal{V}\Lambda\mathcal{V}^\dagger\mathcal{V}\Lambda\mathcal{V}^\dagger}{\mathrm{Tr}[\mathcal{V}\Lambda\mathcal{V}^\dagger\mathcal{V}\Lambda\mathcal{V}^\dagger]} = \mathcal{V}\frac{\Lambda^2}{\mathrm{Tr}[\Lambda^2]}\mathcal{V}^\dagger = \mathcal{V}\Sigma\mathcal{V}^\dagger, \tag{14}$$

where $T$ admits the spectral decomposition $\mathcal{V}\Lambda\mathcal{V}^\dagger$, and $\Sigma = \Lambda^2/\mathrm{Tr}[\Lambda^2]$ is automatically a non-negative diagonal matrix with unit trace. We remark that the factored method in Eq. (14) corresponds to an eigenvalue mapping $\mathcal{P}(\cdot)$ as $\mathcal{P}(\Lambda) = \Lambda^2/\mathrm{Tr}[\Lambda^2]$, and can also be regarded as a state-mapping strategy which first uses an eigenvalue mapping (e.g., $\mathcal{P}(\Lambda) = |\Lambda|$) to process the Hermitian $T$ and simultaneously uses the state factorization to output a physical state $\tilde{\rho}$, in the form of

$$\tilde{\rho} = \frac{\mathcal{P}[T^\dagger]\mathcal{P}[T]}{\mathrm{Tr}[\mathcal{P}[T^\dagger]\mathcal{P}[T]]} = \mathcal{V}\frac{\mathcal{P}(\Lambda)^2}{\mathrm{Tr}[\mathcal{P}(\Lambda)^2]}\mathcal{V}^\dagger. \tag{15}$$

Here the eigenvalue mapping $\mathcal{P}(\cdot)^2$ contains the projection in Sec. II C 2, without the need to map eigenvalues to the unit simplex.

Following further from Eq. (15), we propose the $P$-order absolute state mapping,

$$\tilde{\rho} = \mathcal{A}_P[T] = \mathcal{V}\frac{|\Lambda|^P}{\mathrm{Tr}[|\Lambda|^P]}\mathcal{V}^\dagger, \tag{16}$$

with the eigenvalue mapping

$$\mathcal{P}_P(\Lambda) = |\Lambda|^{\frac{P}{2}}. \tag{17}$$

The tunable parameter $P$ is introduced to adjust the weighting of different eigenvalues of $\Lambda$. It is easy to verify that Eq. (14) is a special case of $\mathcal{A}_P[\cdot]$ with $P = 2$, and $\mathcal{A}_P[\cdot]$ with an even integer $P$ can be efficiently computed similar to Eq. (14), using only matrix multiplication rather than costly spectral decomposition.

The parameter $P$ enables the $\mathcal{A}_P$ mapping in Eq. (16) to accommodate diverse tomography scenarios, addressing the requirements for both full-rank and rank-deficient estimation. In terms of computational efficiency and tomography accuracy, $P$ is recommended as 2 for the general unknown states and choosing a larger even-integer $P$ for the rank-deficient estimation, as supposed by the experimental findings in Sec. IV.

#### 2. Rationality analysis

We show that for an arbitrary density matrix $\tilde{\rho}$, there always exists at least one Hermitian matrix $T$ such that $\tilde{\rho} = \mathcal{A}_P[T]$.

*Theorem 1.* Let $\tilde{\rho}$ be an arbitrary physical density matrix and $\mathcal{A}_P[\cdot]$ the state mapping $\mathcal{A}_P[T] : T \to \tilde{\rho} = \mathcal{V}(|\Lambda|^P/\mathrm{Tr}[|\Lambda|^P])\mathcal{V}^\dagger$ for Hermitian matrix $T$ with spectral decomposition $\mathcal{V}\Lambda\mathcal{V}^\dagger$. Then, there always exists a surjection mapping $\mathcal{A}_P[\cdot]$.

The proof is sketched as follows. Note that $\tilde{\rho}$ obeys the decomposition $\tilde{\rho} = \sum_i p_i|\psi_i\rangle\langle\psi_i|$ in Eq. (1). By setting $T = \sum_i (p_i)^{1/P}|\psi_i\rangle\langle\psi_i|$, we have $\tilde{\rho} = \mathcal{A}_P[T]$. A similar construction is also obtained as $\sum_i -(p_i)^{1/P}|\psi_i\rangle\langle\psi_i|$. Thus, the mapping $\mathcal{A}_P[\cdot]$ exists and is surjection.

The proof of Theorem 1 indicates that the eigenvalues of $\tilde{\rho}$ and matrix $T$ are in one-to-one correspondence, allowing for iteratively optimizing eigenvalues of $\tilde{\rho}$. Moreover, the $\mathcal{A}_P$ mapping preserves the full eigenvalues information without discarding the negative ones, improving the tomography performance and also alleviating the rank-deficient issue for full-rank states with suitable $P$, to be confirmed by the experiments reported in Sec. IV. We remind the reader that the excellent fit of our method to the measured data may result in overfitting for large noisy data, which can be mitigated by performing rank-deficient estimation using a larger $P$. It is also possible to apply $\mathcal{A}_P$ to known QST estimators, such as the MLE with MGD [41], to provide accuracy and convergence improvements.

We remark that there exists a way to unite the factored and projected methods in which the factored method is used in the early stage to achieve a fast convergence and then the projected method is employed to improve final tomography accuracy, such as the MLE with CG-APG [40]. However, our numerical results in Sec. IV C show that the MLE with CG-APG becomes inefficient to tomography for the large-qubit states due to the CG optimizer. Besides, there are issues in CG-APG regarding the switching condition between the factored and projected methods as well as the computational cost to be carefully examined.

### B. Momentum-accelerated Rprop gradient algorithm

The MRprop gradient algorithm is then proposed to accelerate the state search, where model parameters are updated based on the sign of gradients, the step size depends on two consecutive changes in the gradients, and a consistent direction enlarges the step size and, vice versa, decreases it [53]. We further introduce the gradient momentum to ensure more stable and faster convergence, and the complete MRprop gradient algorithm with $\mathcal{A}_P$ mapping is summarized as Algorithm I.

In addition to gradient momentum, we also add judgment on the change of objective function. If the gradient change is opposite and the function value is greater than the last one, then the gradient direction is inverted, which further speeds up the convergence of the algorithm. The comparison of our developed method with other QST algorithms is showcased in Sec. IV C.

### C. POVM with product structure

In this work, the $n$-qubit POVM $\{M_{\mathbf{k}}\}$ is chosen as a tensor product of single-qubit POVM $\{M_k\}_{k=1}^K$, where $\mathbf{k} = (k_1, \dots, k_n)$ with $k_i \in \{1, \dots, K\}$ and $M_{\mathbf{k}} = M_{k_1} \otimes \cdots \otimes M_{k_n}$.

ALGORITHM 1: MRprop gradient algorithm with $\mathcal{A}_P$-mapping.

**Input:** $\{f_k\}$, $\{M_k\}$, $\boldsymbol{\theta}_0 \in \mathbb{R}^{d^2}$, $\eta_+ = 1.14$, $\eta_- = 0.66$, $\tau_{\max} = 50$, $\tau_{\min} = 10^{-6}$, $\mu = 10^{-2}$
**Output:** $\tilde{\rho}(\theta_t)$

Initialize $\boldsymbol{\theta}_0$ randomly, $\eta_0 = 0.5$, $\boldsymbol{m}_0 = \mathbf{0}$;
**for** $t \leftarrow 1, \ldots,$ **do**
  $\tilde{\rho}_t(\boldsymbol{\theta}_{t-1}) \leftarrow \mathcal{A}_P[T(\boldsymbol{\theta}_{t-1})]$;
  $l_t(\boldsymbol{\theta}_{t-1}) \leftarrow -\sum_k f_k \ln(\mathrm{Tr}\,[M_k \tilde{\rho}_t(\boldsymbol{\theta}_{t-1})])$;
  $\boldsymbol{g}_t \leftarrow \nabla_{\boldsymbol{\theta}} l_t(\boldsymbol{\theta}_{t-1})$;
  **for** $i \leftarrow 0$ **to** $d^2 - 1$ **do**
    **if** $g^i_{t-1} g^i_t > 0$ **then**
      $\eta^i_t \leftarrow \min\{\eta^i_{t-1}\eta_+, \tau_{\max}\}$;
    **else**
      **if** $g^i_{t-1} g^i_t < 0$ **then**
        $\eta^i_t \leftarrow \max\{\eta^i_{t-1}\eta_-, \tau_{\min}\}$;
        **if** $l_t > l_{t-1}$ **then**
          $g^i_t \leftarrow -g^i_t$;
        **else**
          $g^i_t \leftarrow 0$;
        **end**
      **else**
        $\eta^i_t \leftarrow \eta^i_{t-1}$;
      **end**
    **end**
  **end**
  **if** $\mu > 0$ **then**
    $\boldsymbol{m}_t \leftarrow \mu \cdot \boldsymbol{m}_{t-1} + \eta_t \cdot \mathrm{sign}(\boldsymbol{g}_t)$;
    $\boldsymbol{\theta}_t \leftarrow \boldsymbol{\theta}_{t-1} - \boldsymbol{m}_t$;
  **else**
    $\boldsymbol{\theta}_t \leftarrow \boldsymbol{\theta}_{t-1} - \eta_t \cdot \mathrm{sign}(\boldsymbol{g}_t)$;
  **end**
  $\boldsymbol{g}_{t-1} \leftarrow \boldsymbol{g}_t$;
**end**

The total number of measurement settings $\{M_{\mathbf{k}}\}$ is $K^n$, thus yielding a $K^n$-dimensional PD. Consequently, obtaining $P_{\mathbf{k}} = \mathrm{Tr}[M_{\mathbf{k}}\rho]$ reduces to the product of the matrices.

We take the two-qubit state as an example. Given matrices $M_i = [\begin{smallmatrix} a_i & b_i \\ c_i & d_i \end{smallmatrix}]$ and $\rho = [\begin{smallmatrix} A & B \\ C & D \end{smallmatrix}]$, where $A$, $B$, $C$, $D$ are block matrix elements, there is

$$\begin{aligned} P_{ij} &= \mathrm{Tr}[(M_i \otimes M_j)\rho] \\ &= \mathrm{Tr}\left[(M_i \otimes M_j)\left(\begin{bmatrix} 1 & 0 \\ 0 & 0 \end{bmatrix} \otimes A + \begin{bmatrix} 0 & 1 \\ 0 & 0 \end{bmatrix} \otimes B + \cdots\right)\right] \\ &= a_i \mathrm{Tr}[M_j A] + b_i \mathrm{Tr}[M_j B] + c_i \mathrm{Tr}[M_j C] + d_i \mathrm{Tr}[M_j D] \\ &= V_{M_i}^{\mathsf{T}} [V_A \quad V_B \quad V_C \quad V_D]^{\mathsf{T}} V_{M_j}, \end{aligned} \tag{18}$$

where $V_X$ is the column vectorization of matrix $X$, such as $V_{M_i} = [a_i\ b_i\ c_i\ d_i]^{\mathsf{T}}$. This form can greatly reduce the computational and storage costs, which has been verified in [40], where the cost of computing the probabilities is reduced from $O(K^n 4^n)$ to $O(K^{n+1})$. Thus, we utilize the product-structured POVM in all QST algorithms tested in the following section.

## IV. NUMERICAL EXPERIMENTS

In this section, we perform extensive experiments to verify the great utility of the $\mathcal{A}_P$-mapping method in various tomography scenarios. In particular, it is tested on two large classes of states. One is a class of $n$-qubit Werner states,

$$\rho = p|\psi\rangle\langle\psi| + \frac{1-p}{d}\mathbb{1}, \tag{19}$$

with $0 \leqslant p \leqslant 1$ and $d = 2^n$. The pure state $|\psi\rangle$ is randomly generated from

$$|\Psi\rangle = \frac{\sum_{k=0}^{d-1} \Psi_k |k\rangle}{\sqrt{\sum_{k=0}^{d-1} |\Psi_k|^2}}, \tag{20}$$

where $\Psi_k$ are sampled from $\{0, 1, -1, i, -i\}$ and $|k\rangle$ denotes the state basis with binary $k$ (e.g., $|2\rangle$ is $|10\rangle$, for $n = 2$). It is obvious that the state set $\{|\Psi\rangle\}$ covers the product, Greenberger-Horne-Zeilinger (GHZ), and W states. The other is a family of mixed states with rank $r$,

$$\rho = \frac{A^\dagger A}{\mathrm{Tr}[A^\dagger A]}, \tag{21}$$

where matrix element $A_{ij}$ in $A \in \mathbb{C}^{r\times d}$ is randomly sampled from $\{0, 1, -1, i, -i\}$ satisfying that $\rho$ has rank $r$.

Numerical experiments are then performed to study the following problems:

(P1) How do the techniques of state mapping affect the tomography performance of the MLE with MRprop and even other QST algorithms?

(P2) What is the reason for the improvements of our $\mathcal{A}_P$-mapping method?

(P3) Is the MLE with MRprop and $\mathcal{A}_P[\cdot]$ capable of efficiently reconstructing various multiqubit states, ranging from pure states sampled from the set $\{|\Psi\rangle\}$ to more general mixed states?

(P4) Are there advantages over previous QST methods with different optimization algorithms, in terms of tomography fidelity, runtime, and iterations, when our MLE with MRprop and $\mathcal{A}_P[\cdot]$ is used?

### A. Experimental setups

The single-qubit tetrahedral POVM [54] is given by $\{M_k = [\mathbb{1}_2 + (\mathbf{s}_k \cdot \boldsymbol{\sigma})/\sqrt{3}]/4\}_{k=1}^4$, with

$$\begin{aligned} &\mathbf{s}_1 = (1, 1, 1),\ \mathbf{s}_2 = (-1, -1, 1), \\ &\mathbf{s}_3 = (-1, 1, -1),\ \mathbf{s}_4 = (1, -1, -1), \end{aligned} \tag{22}$$

where $\boldsymbol{\sigma} = (\sigma_1, \sigma_2, \sigma_3)$. Note that our method can be easily applied to the case involving multiple POVMs on each qubit or other nontetrahedral POVM. To reduce computational cost and easily compare with other QST algorithms, we focus on the tetrahedral POVM. Measurement samples $\{N_{\mathbf{k}}\}$ are acquired via numerically sampling the multinomial distribution $\{p_{\mathbf{k}}\}$ of $N$ times, with $p_{\mathbf{k}} = \mathrm{Tr}[M_{\mathbf{k}}\rho]$ for the $n$-qubit POVM $\{M_{\mathbf{k}}\}$ and target state $\rho$, and then the measurement frequency $f_{\mathbf{k}}$ is obtained as $N_{\mathbf{k}}/N$.

Numerical experiments are conducted on the computer with single Intel(R) Core(TM) i7-12700KF CPU @ 3.60 GHz with 64 GB RAM, and single NVIDIA GeForce RTX 3090Ti

GPU with 24 GB RAM. All QST algorithms are run on the platform of PYTORCH 2.01 and CUDA 11.8 with the same GPU device to speed up computation. The parameters of the MRprop algorithm are shown in Algorithm 1, identical to those used in MGD, and the other QST algorithms are consistent with the original work [12,40,44]. The codes are available in Ref. [55]. In addition to state fidelity (12), we also collect the runtime [12,17,40] and number of iterations [8,28] to evaluate the performance of QST algorithms in Sec. IV C. In detail, the runtime accumulates the minimum time spent per iteration of the algorithm, and one iteration means obtaining an estimate of the target state. Both the runtime and iterations are recorded until the state fidelity reaches 0.99 or the algorithm consumes the maximum allowed iterations.

### B. Performance of the $\mathcal{A}_P$-mapping method

The tomography performance of $\mathcal{A}_P$ mapping is examined in terms of convergence speed, accuracy in various sample number $N$, and accuracy on states with various purity and rank. By contrast, other mapping methods acting on our MLE with MRprop, MLE with MGD [41], and LRE [12], are also investigated. In summary, there is a total of 10 state mappings involved:

(i) the factorization method (6) with lower triangular matrix $\text{Fac}_\text{T}$ and general matrix $\text{Fac}_\text{A}$;

(ii) the $\mathcal{S}$ and $\mathcal{M}$ projection mentioned in Sec. II C 2 for the regular gradient algorithm and projected gradient algorithm (labeled as "proj");

(iii) the factorization method with Hermitian matrix (14), equal to $\mathcal{A}_2$; and

(iv) the $P$-order absolute state mapping $\mathcal{A}_P$ (16) with $P = 1, 3, 4$.

The target states are chosen as eight-qubit mixed states that contain 11 Werner states (19) with $p = \sqrt{(\gamma - 1/d)/(1 - 1/d)}$ and state purity $\gamma$ sampled equidistant from the interval $[1/d, 0.99]$ and nine random mixed states (21) with varying rank $r$, which is sampled via $\log_2 r$ equidistant from $[0, 8]$. The number of iterations is $10^3$ and $5 \times 10^3$ for the MLE with MRprop and MLE with MGD, respectively, and infidelity is compared after reaching the value of $10^{-4}$ or all iterations.

#### 1. *Tomography convergence speed*

We first examine the tomography infidelity convergence of our MLE with MRprop based on 10 mappings with $N = 10^{10}$ measurement samples. It is shown in Fig. 1 that our $\text{Fac}_\text{H}$ achieves the fastest convergence and the best tomography accuracy, with state infidelity of almost $10^{-3}$ within $10^3$ iterations. It is also found that other factored methods $\text{Fac}_\text{A}$ and $\text{Fac}_\text{T}$ quickly converge at the early stage, but perform worse than $\text{Fac}_\text{H}$ by up to an order of magnitude in accuracy at the final stage. The $\mathcal{S}$ and $\mathcal{M}$ projections offer no convergence advantage for a regular gradient algorithm, whereas they admit a rapid convergence in the later stage for the projected gradient algorithm.

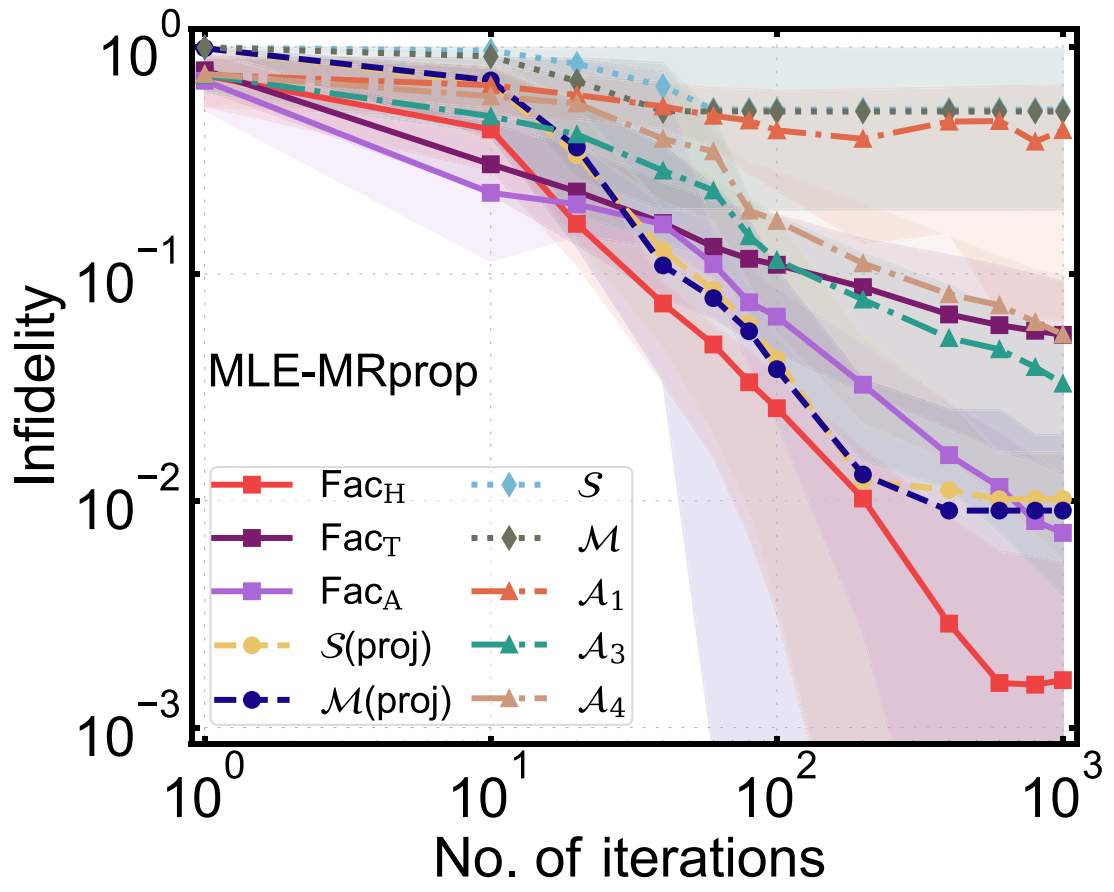

FIG. 1. The tomography infidelity convergence of our MLE with MRprop based on 10 state mappings, with $N = 10^{10}$ measurement samples. Each is tested on 20 states that contain 11 random Werner states (19) and nine random mixed states (21). The line denotes median infidelity and the shaded area is the interquartile range around the median.

#### 2. *Tomography accuracy with varying sample*

We then implement the mapping experiments with measurement samples $N = 10^7, 10^8, 10^9, 10^{10}, 10^{11}$ for our MLE with MRprop, MLE with MGD, and LRE. As shown in Fig. 2(a), $\text{Fac}_\text{H}$ outperforms other mapping techniques for MLE with MRprop, such that more than an order of magnitude improvement in infidelity is obtained. Importantly, it admits a $O(1/N)$ infidelity scaling as the infidelity scales from $10^{-1}$ to $10^{-4}$ as the sample number $N$ varies from $10^8$ to $10^{11}$. As further displayed in Figs. 2(b) and 2(c), the tomography accuracy of MLE with MGD is similar to that of MLE with MRprop, indicating that the $\mathcal{A}_P$ mapping is widely applicable to various optimizers. For LRE, $\mathcal{S}$ and $\mathcal{M}$ projections have a better infidelity but a worse tomography accuracy, in comparison with the MLE. This matches well with the previous work [11].

#### 3. *Tomography accuracy on states with varying purity*

We continue to study how these mapping methods act on full-rank states with different purity $\gamma$. Similar to the experimental settings in Sec. IV B 2, we test the MLE with MRprop, MLE with MGD, and LRE based on 10 state mappings by reconstructing 11 full-rank Werner states (19) with uniformly distributed purity.

In Fig. 3(a), there is a trend that the increase of $P$ in the MLE with MRprop leads to a high state purity, for the $\mathcal{A}_p$ mapping. For $P = 2$, $\text{Fac}_\text{H}$ achieves the best accuracy on states across purity. Similar results are obtained for the case of MLE with MGD in Fig. 3(b), except that $\mathcal{A}_1$ also demonstrates good performance. It is further found in Fig. 3(c) that for the LRE, $\mathcal{A}_1$ has comparable accuracy as $\mathcal{S}$ and $\mathcal{M}$ on most tested states.

#### 4. *Tomography accuracy on states with varying rank*

We further perform a full-state tomography of nine random mixed states (21) with some chosen ranks $r$ based on the above 10 state mappings. The results in Fig. 4(a) show that for the MLE with MRprop, $\mathrm{Fac_H}$ has the highest fidelity on states with various rank, while $\mathrm{Fac_A}$ and the projected gradient algorithms based on $\mathcal{S}$ and $\mathcal{M}$ attain similar accuracy on most states. However, $\mathcal{A}_P$ with a large $P$ tends to suffer from rank-deficient estimation, which is explained in Sec. IV B 5. Similar results are obtained for MLE with MGD in Fig. 4(b). These results suggest that $\mathrm{Fac_H}$ offers the strongest adaptation to a wide range of states. Several good matches are established between the $\mathcal{S}$ and $\mathcal{M}$ projections and $\mathrm{Fac_A}$, $\mathcal{A}_P$ with large $P$ and $\mathrm{Fac_T}$, in the case of MLE.

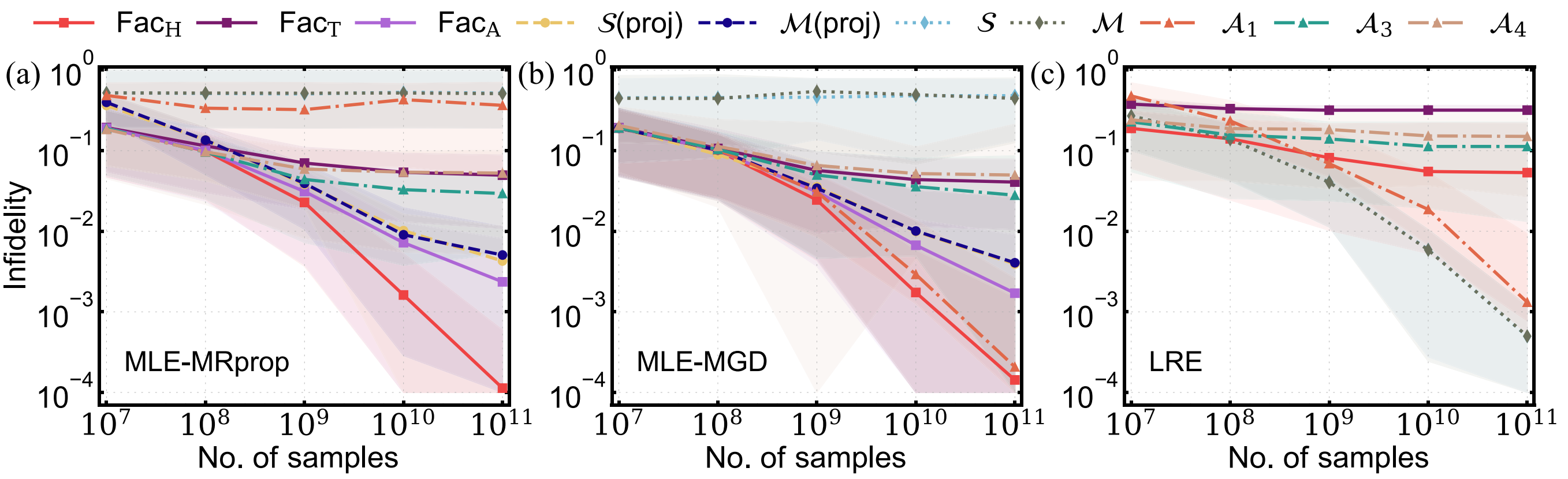

FIG. 2. The tomography infidelity of (a) our MLE with MRprop, (b) MLE with MGD, and (c) LRE based on 10 mappings, with the sample size of $N = 10^7, 10^8, 10^9, 10^{10}, 10^{11}$. The median infidelity obtained from 20 experiments with 11 Werner states (19) and nine random mixed states (21) is described as the curves, with the interquartile range as the shaded area.

#### *5. Analysis of performance improvements*

We finally study the problem of whether the eigenvalues of density matrices reconstructed by $\mathcal{A}_P$ match with those of the target states. Specifically, the target states are generated as eight-qubit Werner states (19) with purity $\gamma = 0.004, 0.596$, and random mixed states (21) with rank $r = 32, 256$. Without loss of generality, we examine this match problem via MLE with MRprop under nine mappings and $10^{10}$ measurement samples. The eight largest and eight smallest eigenvalues are picked up to benchmark the corresponding algorithm.

As illustrated in Fig. 5, it is observed that $\mathrm{Fac_H}$ fits the actual eigenvalues almost perfectly for states with various purity and rank, demonstrating its strong adaptability for a wide range of states. For example, the minimum eigenvalue of the Werner state with $\gamma = 0.004$ is $3.9 \times 10^{-3}$, which is estimated as $3.5 \times 10^{-3}$ by $\mathrm{Fac_H}$, while estimates of $\mathrm{Fac_A}$, $\mathcal{A}_4$, and $\mathcal{S}$ and $\mathcal{M}$ used for the projected gradient are $3.1 \times 10^{-3}$, $3.9 \times 10^{-10}$, $1.1 \times 10^{-3}$, $2.2 \times 10^{-3}$, respectively. For the mixed state with $r = 256$, the actual and estimate minima are $2.9 \times 10^{-7}$, $8.9 \times 10^{-7}$, $3.4 \times 10^{-8}$, 0, 0, 0, respectively. These results confirm that $\mathrm{Fac_H}$, equivalent to $\mathcal{A}_P$ mapping with $P = 2$, is suitable for the MLE, with excellent accuracy and efficient computation.

### C. Performance of various QST algorithms

We then introduce tomography infidelity, runtime, and number of iterations as figures of merit, to evaluate the tomography performance of various QST algorithms on random Werner states within $10^4$ iterations. Particularly, these algorithms are our MLE with MRprop and $\mathrm{Fac}_\mathrm{H}$ mapping, MLE with MGD [41] and $\mathrm{Fac}_\mathrm{H}$ mapping, iMLE [44], MLE with CG-APG [40] using $\mathrm{Fac}_\mathrm{A}$ mapping and $\mathcal{S}$ projection, and LRE [12] with $\mathcal{S}$ projection and $\mathcal{A}_1$ mapping.

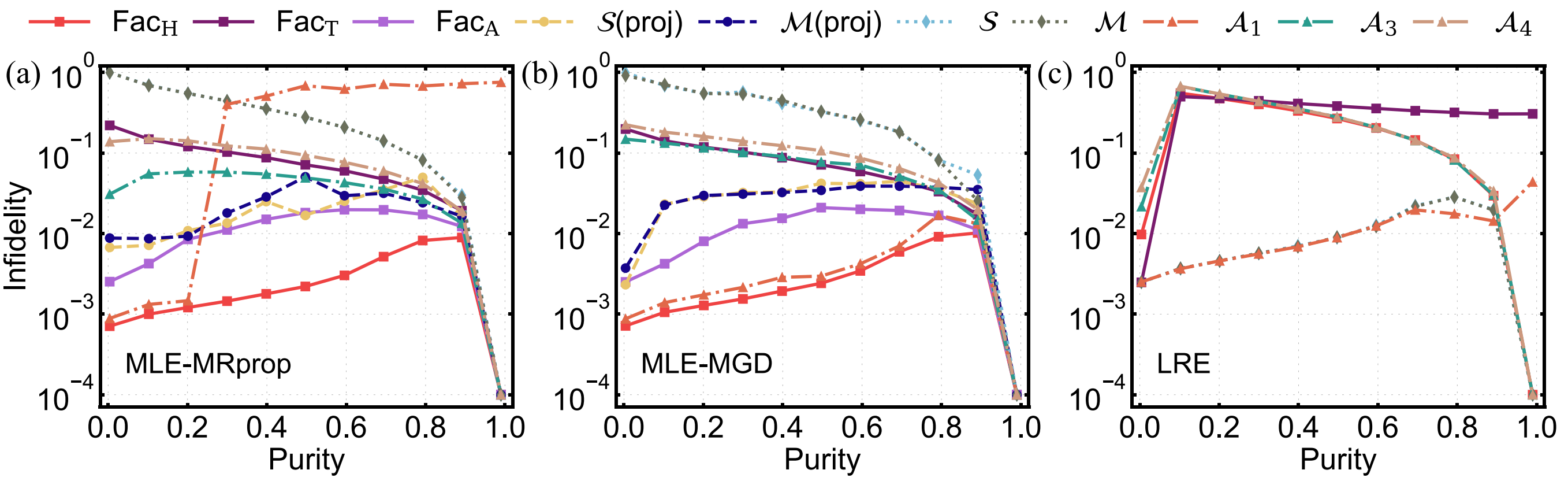

FIG. 3. The tomography infidelity of (a) our MLE with MRprop, (b) MLE with MGD, and (c) LRE for Werner states (19) with uniformly distributed purity, based on the 10 mappings with $N = 10^{10}$ samples. The 11 full-rank Werner states are chosen with $p = \sqrt{(\gamma - 1/d)/(1 - 1/d)}$ and $\gamma$ sampled from $1/d$ to 0.99.

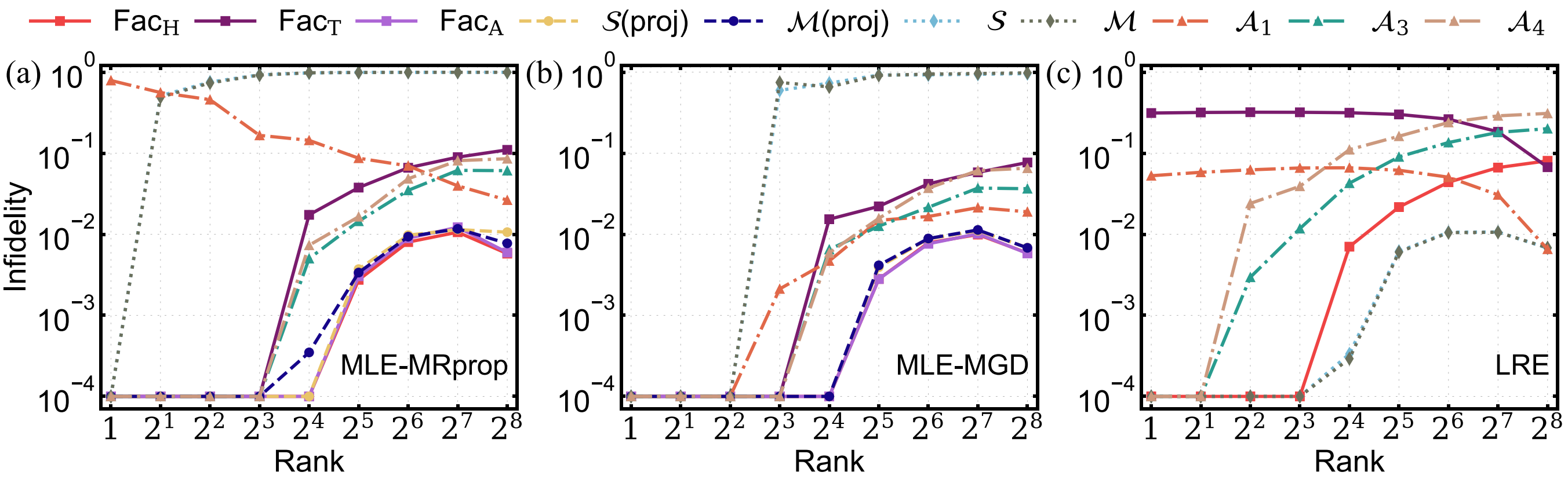

FIG. 4. The tomography infidelity of (a) our MLE with MRprop, (b) MLE with MGD, and (c) LRE for random mixed states (21) with various rank, based on the 10 mappings with $N = 10^{10}$ samples. The nine mixed states are chosen with $\log_2 r$ sampled equidistant from 0 to 8.

### *1. Tomography of eight-qubit states with varying sample*

First, these QST algorithms are tested via the task of reconstructing the randomly generated eight-qubit Werner states, subjected to the samples noise. It is shown in Fig. 6(a) that the MLE with MRprop or MGD attains a higher fidelity than others, with approximately an order of magnitude improvement in state fidelity. Furthermore, it is found in Fig. 6(b) that our MLE with MRprop consumes fewer iterations and runtime, compared to the MLE with both MGD and CG-APG. The iMLE has very slow convergence and does not always reach 0.99 fidelity even with $10^4$ iterations. The LRE requires no iteration and is extremely fast, making it suitable for yielding an initial estimate of other QST algorithms.

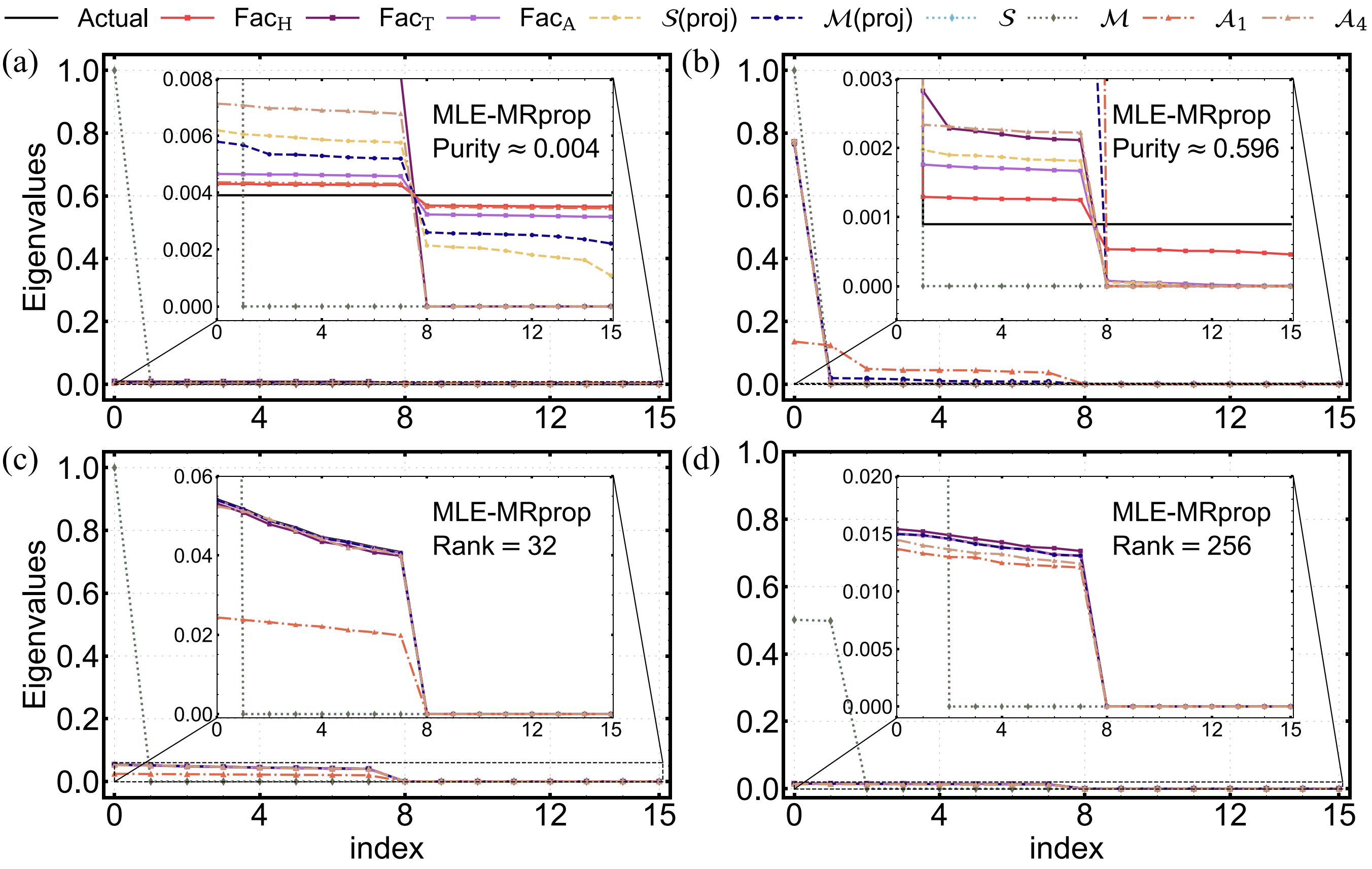

FIG. 5. The distribution of actual and estimate eigenvalues of our MLE with MRprop for the eight-qubit Werner states (19) with state purity (a) 0.004 and (b) 0.596, as well as random mixed states (21) with rank (c) 32 and (d) 256, within $10^{10}$ measurement samples. The actual and estimate 16 eigenvalues are selected as the eight largest and the eight smallest of the density matrix in descending order.

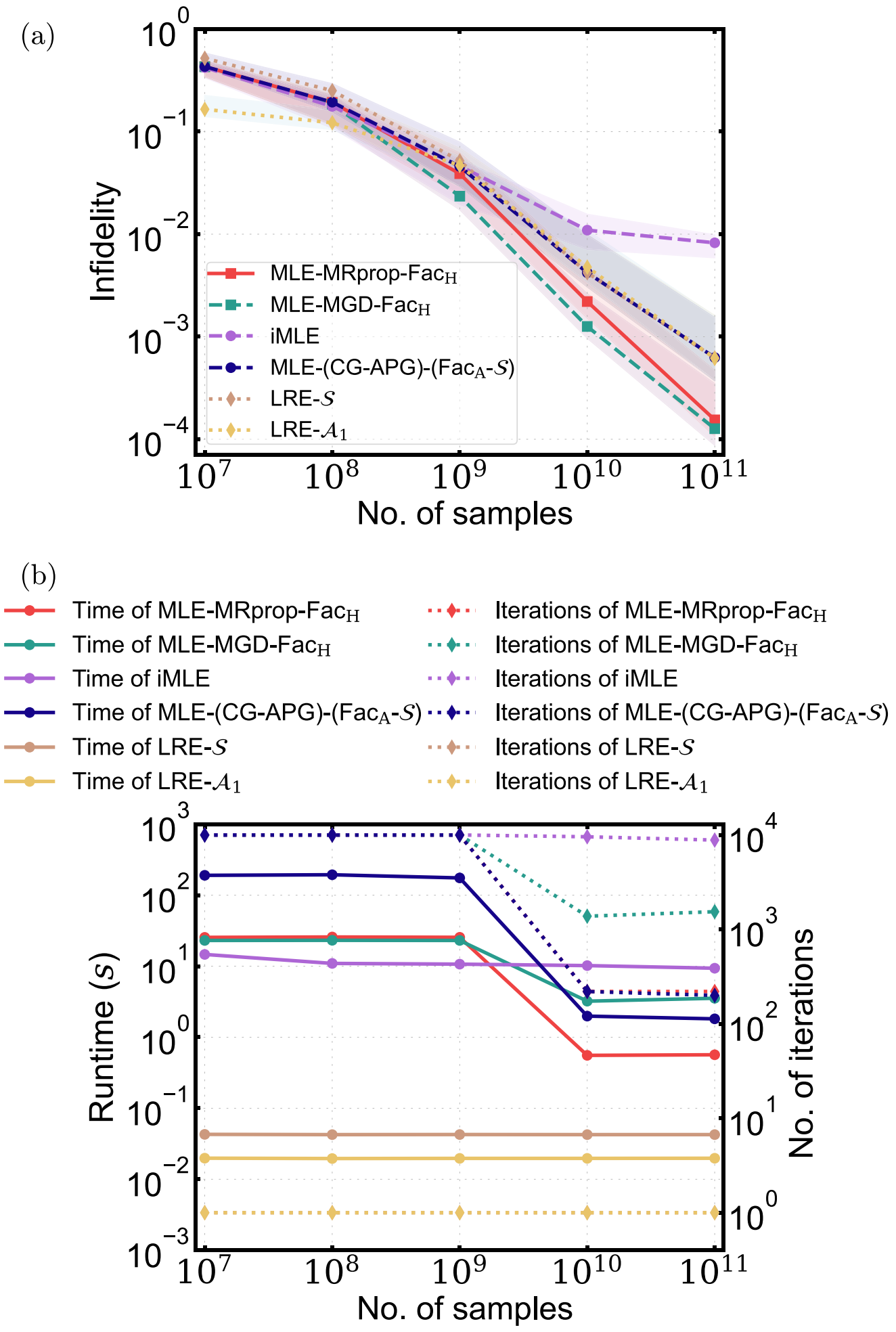

FIG. 6. The (a) tomography infidelity and (b) runtime and iteration number of our MLE with MRprop, MLE with MGD, iMLE, MLE with CG-APG, and two LREs. Given the sample size of $N = 10^7, 10^8, 10^9, 10^{10}, 10^{11}$, each experiment is repeated 20 times with the eight-qubit random Werner states. Each data in (a) and (b) represents the median metric and the shaded area in (a) is the interquartile range.

### 2. *Tomography of states up to 11 qubits*

We further extend the eight-qubit experiments to states up to 11 qubits, accounting for runtime and number of iterations. The maximum number of iterations is set to $10^4$ and statistical noise is not considered.

It follows from Fig. 7 that our MLE with MRprop and $\mathrm{Fac_H}$ mapping is able to precisely tomography 11-qubit states within one minute, implying the strong scalability of our method. It outperforms the MLE with CG-APG and the iMLE in that it takes one or two orders of magnitude less runtime than the MLE with CG-APG for 9- and 10-qubit states and the iMLE for states with >6 qubits. Further, our MLE with MRprop shows excellent scalability on a large number of qubits (>8), with less number of iterations and runtime than the MLE with the CG-APG algorithm. The iMLE is eliminated due to slow convergence. Finally, it is worth noting that although the LRE reaches the same fidelity faster than our MLE with MRprop, it results in a lower-fidelity estimate when the same measurement samples are used, as detailed in Sec. IV C 1.

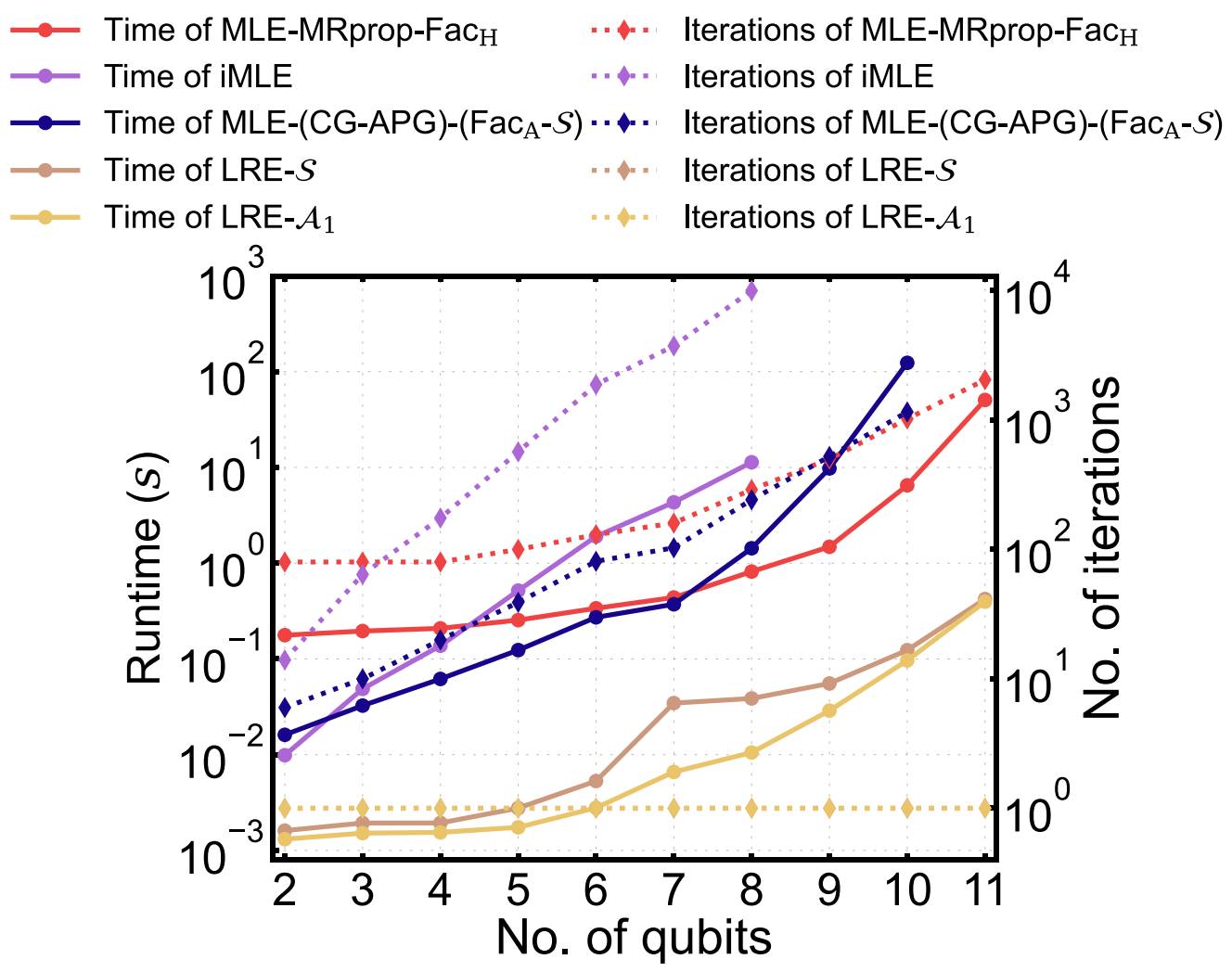

FIG. 7. The runtime and iteration number of our MLE with MRprop and $\mathrm{Fac_H}$ mapping and other QST algorithms with the varying number of qubits. Experiments are implemented to reconstruct Werner states with uniformly distributed $p$. A data point is obtained by mediating either the time or iteration number over 20 states. The iMLE is up to eight-qubit states as it cannot achieve the 0.99 fidelity within $10^4$ iterations, while the MLE with CG-APG is up to 10 qubits limited by long runtime.

## V. CONCLUSION AND OUTLOOK

We have presented the factored gradient descent algorithm which combines a $P$-order absolute state-mapping technique (16) with a momentum-accelerated Rprop gradient algorithm to achieve fast and accurate reconstruction of multiqubit states. With extensive experiments on the proposed state-mapping strategy and other QST algorithms, sufficient results show that our scheme effectively fits the eigenvalues and achieves strong adaptability to the state purity and rank, as well as more precise tomography in less runtime and iterations. It is further demonstrated that the our MLE method enables tomography of 11-qubit mixed states in less than one minute and achieves optimal $O(1/N)$ infidelity scaling with sample size $N$. Finally, our state-mapping strategy also provides the performance improvement for other QST algorithms.

There are many interesting questions left for future work. For instance, a suitable construction of the estimate matrix in a factored method is still needed to address the rank-deficient issue, and the discovery of other non-Hermitian constructions can deepen the understanding of state mappings. Further, it would also be interesting to verify our method on a real quantum chip.

## ACKNOWLEDGMENTS

We thank Dr. C. Hu for useful discussions and also thank the anonymous referee for the constructive comments. This research is supported by the National Natural

Science Foundation of China (Grants No. 12205219, No. 61903027, No. 72171172, No. 62088101); the National Key R&D Program of China (Grants No. 2018YFE0105000, No. 2018YFB1305304); the Shanghai Municipal Science and Technology Major Project (Grant No. 2021SHZDZX0100); the Shanghai Municipal Commission of Science and Technology (Grants No. 1951113210, No. 19511132101); the Shanghai Research Institute of China Engineering Science and Technology Development Strategy, Strategic Research and Consulting Project (Grant No. 2022-DFZD-33-02); the Shanghai Municipal Science and Technology Fundamental Project (Grant No. 21JC1405400); and the Chinese Academy of Engineering, Strategic Research and Consulting Program (Grant No. 2022-XY-100).